\documentclass[%
 amsmath,amssymb,
 aps,
 pre,
twocolumn,
 superscriptaddress,
]{revtex4-2}

\usepackage{graphicx}
\usepackage{dcolumn}
\usepackage{bm}

\begin{document}

\title{Diffusion through complex confining environments: fluctuating triply periodic minimal surfaces}

\author{Jakob Mihatsch}
\email{jakob.mihatsch@ovgu.de}
\affiliation{Institut f{\"u}r Physik, Otto-von-Guericke-Universit{\"a}t Magdeburg, Universit{\"a}tsplatz 2, D-39106 Magdeburg, Germany}

\author{Andreas M. Menzel}
\email{a.menzel@ovgu.de}
\affiliation{Institut f{\"u}r Physik, Otto-von-Guericke-Universit{\"a}t Magdeburg, Universit{\"a}tsplatz 2, D-39106 Magdeburg, Germany}

\begin{abstract}
The transport of individual entities through interconnected structures is a process of practical relevance both in biology and technology. Examples are given by diffusive dynamics of molecules in porous structures.
In soft environments, this transport can be strongly influenced by fluctuations of the porous structure itself.
Here, we focus on triply periodic membrane structures found both in cell organelles and in synthetic amphiphilic systems.
We theoretically study the effect of a complex three-dimensional fluctuating environment on the diffusive motion of a test object, using a phase field approach.
The rigid spherical test object is energetically forced to not penetrate the membrane. Generally, the pores of the membrane structure can be smaller than the diffusing object.
Yet, fluctuations of the membrane can intermittently widen its pores, still allowing for the motion of the larger particles through them.
Thus, the object stays trapped for a while inside one cavity formed by the membrane, before an appropriate fluctuation event widens a membrane pore in the right moment so that the object can jump into the next cavity. 
The process is reflected by a pronounced plateau in the time evolution of the mean squared displacement.
We think that the described scenario should be directly observable, for instance, in protein diffusion through biological environments.
\end{abstract}

\maketitle
\section{Introduction}
The diffusive transport of molecules is of interest in biophysics and beyond.
Often, this transport happens in porous, heterogeneous environments.
Transport processes in biological cells take place in crowded environments~\cite{hofling2013anomalous}, and diffusion through nanopores has technological applications such as filtration via membranes or in catalysts~\cite{wu2020nanoparticle}.
Thermal fluctuations do, however, not only lead to the diffusive motion of a molecule, but can also dynamically change the confining environment of this molecule.
There have been experimental \cite{sarfati_enhanced_2021}, numerical~\cite{liNonGaussianNormalDiffusion2019,mohammadi_enhanced_2025}, and analytical~\cite{marbach_transport_2018,palmieri2012diffusion,reister-gottfried_diffusing_2010} efforts to characterize the effect of a fluctuating environment on a diffusing particle.
Depending on the specific setup and the theoretical assumptions, some authors found that the fluctuations of the environment decrease the effective diffusion coefficient~\cite{palmieri2012diffusion,reister-gottfried_diffusing_2010}, others found that it was increased~\cite{sarfati_enhanced_2021,liNonGaussianNormalDiffusion2019,mohammadi_enhanced_2025}.
The latter was especially the case when trapping of the diffusing particle in cavities of the porous medium occurred.
Fluctuations in a heterogeneus environment can be the cause for non-Gaussian diffusion~\cite{liNonGaussianNormalDiffusion2019}. 
There, the mean-squared displacement (MSD) increases linearly in time, while the displacements do not follow a normal distribution.
In many studies, also those mentioned above, the movement is usually limited to a effectively one- or two-dimensional geometry.

Here, we study diffusion in a three-dimensional (3D) fluctuating environment.
Examples of porous structures that inherently exist only in three dimensions are cubic membranes. 
These are complex 3D periodic structures formed by lipid membranes in cell organelles~\cite{almsherqi_cubic_2006,almsherqiCubicMembranesMissing2009}.
It is also possible to create cubic phases in vitro using amphiphilic molecules.

\begin{figure}
    \includegraphics[width=.35\textwidth]{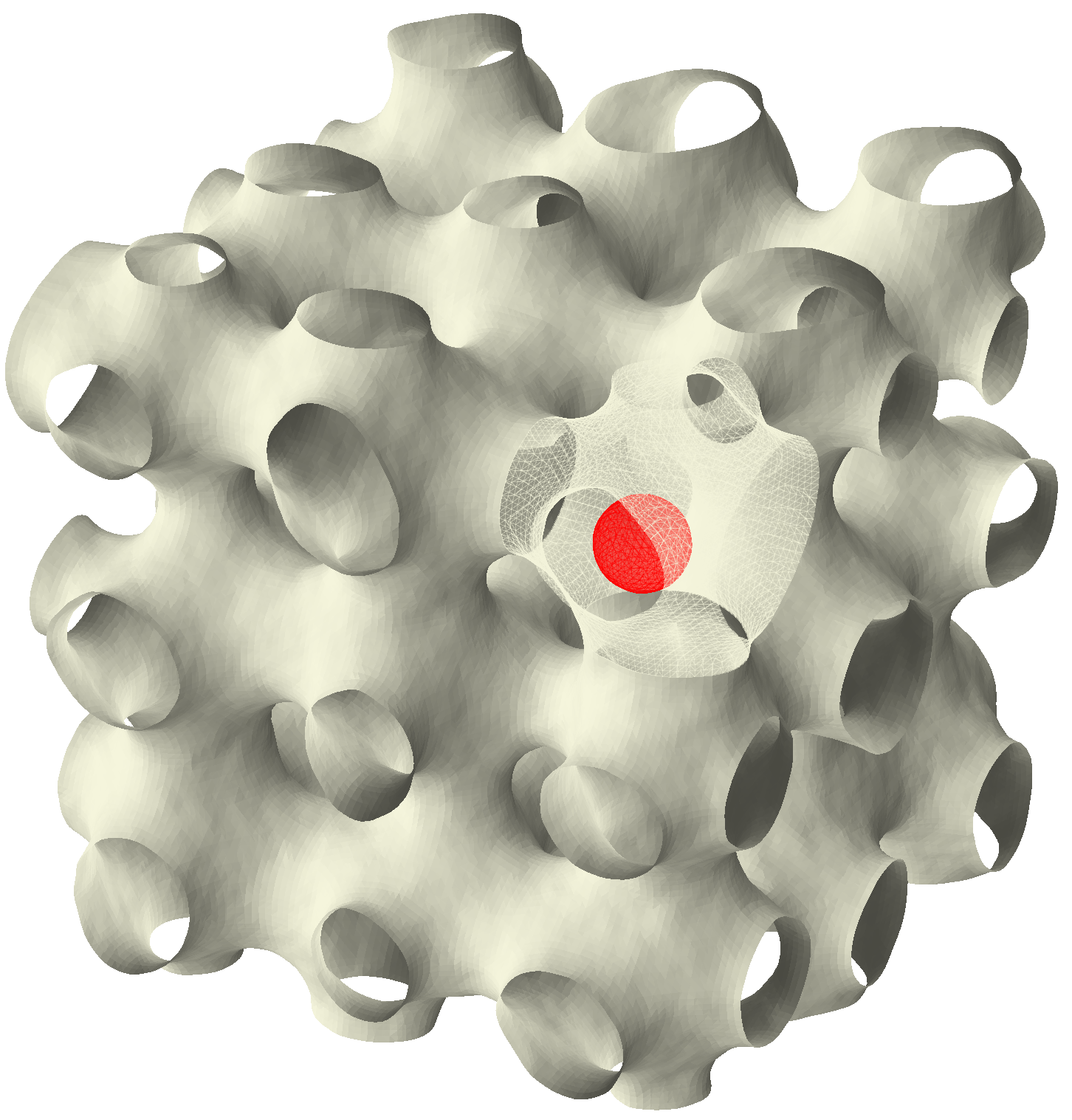}
    \caption{A visualization of the P-surface, of $3\times 3\times 3$ unit cells. The surface separates space into two equal volumes that are intertwined channels.
    A spherical test particle (bright red) can move freely on one side of the surface, but cannot penetrate it. 
    Thermal fluctuations have undulated the surface, so that some pores are open wider than others, and, therefore, at some time, allow the test particle to pass easily. 
    }
    \label{fig:example}
\end{figure}

Mathematically, these structures can be described by periodic surfaces of constant mean curvature (CMC) $H$~\cite{cuiBicontinuousCubicPhases2020}.
In the special case of $H=0$, we observe a so-called triply periodic minimal surface (TPMS).
CMC surfaces are also obtained by minimizing surface area for a given periodicity while prescribing a fixed ratio of the volumes on the two sides of the surface~\cite{grosse-brauckmann_triply_2012}. 
The TPMS is found in the case of equal volume on both sides of the surface.
CMC surfaces can often be thought of as deformed TPMS and are characterized by the TPMS they share their symmetry with.
Important examples of TPMSs are the Schwarz P- and D-surfaces, as well as the gyroid (G-surface) discovered by Schoen~\cite{von_schnering_how_1987,grosse-brauckmann_triply_2012}.
TPMSs and related surfaces separate space into two unconnected volumes that form intertwined channels.
The P-surface can be imagined as a cubic lattice of cavities connected by channels in each coordinate direction.
That means it forms a network of junctions of six different pipes, see Fig.~\ref{fig:example}.
Similarly the D-surface and the gyroid form channels with junctions of four and three pipes, respectively.
Apart from lipid membranes, TPMSs are found in the chitin skeletons of insects, the wings of butterflies, and block-copolymers~\cite{hanOverviewMaterialsTriply2018}.
They also find technological applications, for example in the design of scaffolds for growing tissue~\cite{kapfer2011minimal}.
An exact mechanism behind the formation of those structures in biological systems is not known yet.
The reason why they are energetically favorable is, however, not that they minimize surface area, as lipid membranes do not have a significant surface tension.
Instead, the shape of the membrane molecules dictates a spontaneous curvature $c_0$ that is energetically optimal~\cite{grosse-brauckmann_triply_2012,schroder2006bicontinuous}.
An appropriate energy functional for lipid membranes is therefore the Helfrich free energy~\cite{seifertConfigurationsFluidMembranes1997}
\begin{equation}
    F = \frac{\bar{\kappa}}{2}\int_S (2H-c_0)^2 \mathrm{d}^2r,
    \label{eq:helfrich}
\end{equation}
where the membrane is described by a surface $S$ and has a bending rigidity $\bar{\kappa}$.
A second contribution that depends on an integral of the Gaussian curvature is only relevant if one intends to describe topological changes of the membrane, according to the Gauss-Bonnet theorem~\cite{seifertConfigurationsFluidMembranes1997,grosse-brauckmann_triply_2012}.

Diffusive motion in the periodic network of channels formed by TPMS has already been investigated in the case of static surfaces~\cite{assenzaCurvatureBottlenecksControl2018}.
It was found that bottlenecks, if they are present, have an important influence on the transport properties.
The dynamics of a diffusing particle then may correspond to a hopping process.
It may be trapped in a cavity for a long time before escaping.
In fluorescence correlation spectroscopy measurements of the diffusion coefficient of a protein confined in lipid cubic phases, it was found that the impact of the confinement on the diffusion coefficient was less than that of a comparable confinement with hard walls.
This suggests that the softness of the confinement and thermal fluctuations play an important role~\cite{tanaka2012fluorescence}.
We here investigate how diffusion is influenced when a TPMS is undulated by thermal fluctuations, which will lead to a widening and narrowing of the bottlenecks corresponding to the channels formed by the surface.
An example for the geometry under investigation is shown in Fig.~\ref{fig:example}.
We focus on the P-surface because its structure is easy to visualize and grasp, which facilitates interpretation of the results.
Our approach is, however, straightforward to extend to other CMC surfaces.

\section{The phase-field method}
We use a phase-field approach to track the location of the surfaces.
This method describes a surface as the zero-level set of a continuous scalar function, the phase field $\phi(\mathbf{r})$.
It is obtained by minimizing a free energy $F\left[\phi\right]$, which is a functional of $\phi$.
TPMSs were generated from the phase field model for microemulsions~\cite{gozdzTriplyPeriodicSurfaces1996}, but this model has many free parameters and could result in a profile of the phase field $\phi$ that is difficult to analyze.
Yang et al.~\cite{yangPhasefieldApproachMinimizing2010} showed that by minimizing the free energy associated with the Cahn-Hilliard equation
starting from appropriate initial conditions, one can generate periodic surfaces of constant mean curvature.
We use their approach to generate the initial conditions for our simulations.
This free energy is, however, not suited to simulate the dynamic behavior of those surfaces.
As explained above, the energy should not depend on the surface area of the membrane.
Instead the energy should depend on the mean curvature.
Therefore, we use a free energy developed for the modeling of biological membranes and vesicles~\cite{duPhaseFieldApproach2004,duPhaseFieldCalculus2011,campeloDynamicModelStationary2006,campeloModelCurvaturedrivenPearling2007,lowengrub_phase-field_2009},
\begin{equation}
    F_\mathrm{bend} = \frac{\kappa}{2}\int \mathrm{d}^3r \left[(\phi^2-1)(\phi+\epsilon C_0)-\epsilon^2\nabla^2 \phi\right]^2.
    \label{eq:f_bend}
\end{equation}
This free energy is designed to describe an interface according to the Helfrich energy in Eq.~\eqref{eq:helfrich} with $\bar{\kappa}=2\sqrt{2}\epsilon^3\kappa/3$ and $\sqrt{2}C_0=c_0$~\cite{campeloDynamicModelStationary2006}. 
Consequently, it is minimized by phase fields whose zero-level-set surfaces have constant mean curvature.
One obtains a phase field that has a constant value of $\phi=1$ on one side of the surface and $\phi=-1$ on the other side.
The parameter $\epsilon$ determines the width of the crossover region between these two domains.
For simplicity we focus on minimal surfaces and set $C_0=0$ in the remainder of this article.

The following argument by Campelo et al.~\cite{campeloDynamicModelStationary2006} on why this energy is related to the mean curvature will be useful in our later analysis. 
We make the ansatz that $\phi$ is only a function of the distance $d(\mathbf{r})$ to the interface defined by $\phi=0$.
By writing $\phi=f\left({d(\mathbf{r})}/{\sqrt{2}\epsilon}\right)$, Eq.~\eqref{eq:f_bend} becomes
\begin{equation}
   F_\mathrm{bend} =  \frac{\kappa}{2}\int \mathrm{d}^3r \left[\left(f^{\prime \prime}-\left(f^2-1\right) f\right)+\epsilon f^{\prime} \nabla^2 d(\boldsymbol{r})\right]^2 .
    \end{equation}
Here and in the following, we have left out the argument of $f\left({d(\mathbf{r})}/{\sqrt{2}\epsilon}\right)$ for clarity.
In zeroth order in $\epsilon$, the energy is minimized by $f^{\prime \prime}-\left(f^2-1\right) f=0$, which has the solution 
\begin{equation}
    f(y) = \tanh(y).
    \label{eq:tanh}
\end{equation}
The remaining higher-order term in $F$ is
\begin{equation}
    F_\mathrm{bend} =  \int \mathrm{d}^3r \left[\epsilon f^{\prime} \nabla^2 d(\boldsymbol{r})\right]^2.
    \label{eq:asymptotic}
\end{equation}
It becomes equal to Eq.~\eqref{eq:helfrich} (up to a constant factor, with $c_0=0$) when the integration perpendicular to the surface is carried out, because $f'(z/\sqrt{2}\epsilon)\sim\delta(z)$ in the limit $\epsilon\rightarrow 0$, with $\delta$ denoting the Dirac delta function.

An additional contribution to the free energy arises from the fact that the surface area of the membrane fluctuates around a mean value prescribed by the amount of membrane molecules in the system~\cite{campeloDynamicModelStationary2006,duPhaseFieldApproach2004},
\begin{equation}
    F_\mathrm{A} = \frac{\alpha}{2A_0}\left[A[\phi]-A_0\right]^2,
\end{equation}
where
\begin{equation}
    A[\phi]=\int \mathrm{d}^3r \frac{\epsilon^2}{2}|\nabla \phi |^2.
\end{equation}
This term forces the surface area of the interface to fluctuate around a predefined value set by the parameter $A_0$.
Here, we set $A_0=A[\phi_0]$, where $\phi_0$ is the initial configuration of the phase field.
The value of the parameter $\alpha$ is usually chosen to be very large, so that a constant surface area is enforced.
However, in our case, this would hinder all fluctuations of the surface.
Instead, we choose an intermediate value of $\alpha$, small enough so that the fluctuations on small length scales are dominated by the bending energy in Eq.~\eqref{eq:f_bend},
but large enough so that the overall periodic structure is not destroyed by thermal fluctuations.
Furthermore, the amount of material on both sides of the membrane should be conserved, which leads to a third contribution~\cite{duPhaseFieldApproach2004},
\begin{equation}
    F_\mathrm{V} = \frac{\nu}{2}\left[\left(\int \mathrm{d}^3r\ \phi \right)-V\right]^2.
\end{equation}
The parameter $V$ determines the ratio of volume on the two sides of the interface. For TPMSs $V=0$.
To enforce a sharp value of $V$, we choose an appropriately high value of $\nu$ in our numerics.

We implement the interaction between a diffusing particle and the membrane similar to what had been outlined before~\cite{guSimulatingVesicleSubstrate2014}.
If an object is moving through the porous structure created by the membrane, there is an energetic cost connected to an overlap of this object and the membrane,
\begin{equation}
    F_\mathrm{inter} = \gamma\int \mathrm{d}^3r (\phi(\mathbf{r})^2-1)^2 k\left(\left|\mathbf{r}-\mathbf{R}\right|\right).
\end{equation}
In this expression, $\gamma$ is the strength of the interaction, $\mathbf{R}$ is the center of mass of the object, and $k(r)$ is a kernel function describing its size.
In the following, we choose
\begin{equation}
    k(r) = 1-\tanh\left(\frac{r-l}{\sqrt{2}\epsilon}\right),
    \label{eq:kernel}
\end{equation}
where the parameter $l$ determines the radius of the spherical object.
It should be noted that by implementing the interaction between the object and the membrane in this way, not only will the membrane determine where the particle is allowed to move, but also the presence of the particle will influence the shape of the membrane.

The total free energy $F$ of the system is the sum of all the described contributions,
\begin{equation}
    F = F_\mathrm{bend}+F_\mathrm{A} + F_\mathrm{V} + F_\mathrm{inter}.
\end{equation}
We determine the time evolution of the dynamical variables $\phi$ and $\mathbf{R}$ from
\begin{subequations}
    \begin{eqnarray}
    \frac{\partial\phi(\mathbf{r})}{\partial t} = -M_\phi\frac{\delta F}{\delta \phi} +  \xi_\phi,\\
    \frac{\partial\mathbf{R}}{\partial t} = -M_\mathbf{R}\frac{\partial F}{\partial \mathbf{R}} +  \boldsymbol{\xi}_\mathbf{R},
\end{eqnarray}
    \label{eq:dynamics}
\end{subequations}
where $\langle \xi_\phi(\mathbf{r},t)\xi_\phi(\mathbf{r'},t')\rangle=2M_\phi k_B T\delta(t-t')\delta(\mathbf{r}-\mathbf{r'})$ and $\langle \xi_{\mathbf{R},i}(t)\xi_{\mathbf{R},j}(t')\rangle=2M_\mathbf{R} k_B T\delta(t-t')\delta_{ij}$ for the components $\xi_{\mathbf{R},i},\ i\in\left\{1,2,3\right\}$ of $\boldsymbol{\xi}_{\mathbf{R}}$.
Here $T$ refers to the temperature and $k_B$ is Boltzmann's constant.
It should be noted that our dynamics does not include the fluid that surrounds the membrane and the particle.
These more complicated dynamics are left for future research.

Another option is to use dynamic equations that automatically conserve the integral of $\phi$, eliminating the need for $F_V$~\cite{campeloDynamicModelStationary2006,gallenVesicleFormationInduced2023} (Model B in the nomenclature of Halperin, Hohenberg, and Ma~\cite{halperinRenormalizationgroupMethodsCritical1974}).
They have the disadvantage of producing higher-order spatial derivatives, which makes the numerical approach more involved.
Additionally, the stochastic noise would have to be modified to also conserve volume.
These conservative dynamics likewise do not capture hydrodynamic interactions. Therefore, we use the simpler dynamics described above.
Finally, we note that the (generalized) Fokker-Planck equation for the probability density $P(\phi(\mathbf{r}),\mathbf{R})$, associated with the system of stochastic differential equations, Eqs.~\eqref{eq:dynamics}, has the stationary solution~\cite{karmaPhasefieldModelDendritic1999,halperinRenormalizationgroupMethodsCritical1974}
\begin{equation}
    P(\phi(\mathbf{r}),\mathbf{R})=\frac{1}{Z}\exp\left(-\frac{F}{k_BT}\right),
    \label{eq:prob}
\end{equation}
where
\begin{equation}
    Z = \int \mathrm{D}\phi(\mathbf{r}) \int \mathrm{d}^3R\  \exp\left(-\frac{F}{k_BT}\right),
\end{equation}
and $\mathrm{D}\phi(\mathbf{r})$ denotes functional integration over the phase field.

\section{Membrane fluctuations}
We demonstrate that our description leads to the expected fluctuation spectrum of the membrane, for now ignoring the role of diffusing object in terms of $F_\mathrm{inter}$.
Inserting the asymptotic expression from Eq.~\eqref{eq:asymptotic} into Eq.~\eqref{eq:prob}, we obtain
\begin{equation}
    P(\phi(\mathbf{r}))=\frac{1}{Z}\exp\left(-\frac{\int \mathrm{d}^3r \left[\epsilon f^{\prime} \nabla^2 d(\boldsymbol{r})\right]^2}{k_BT}\right)
    \label{eq:boltzmann}
\end{equation}
For small deviations from the flat surface, the nonlinear terms $F_A$ and $F_V$ can be neglected.
Following Ref.~\onlinecite{karmaPhasefieldModelDendritic1999}, we investigate fluctuations around a flat, rectangular patch of surface $S$.
For this purpose, we introduce the coordinates $z$ perpendicular to the surface and $\mathbf{r}_\perp$ parallel to the surface.
We apply a small perturbation to the surface so that its shape can now be described by the height profile $z=h(\mathbf{r}_\perp)$.
The distance to this perturbed surface is given by
\begin{equation}
    d(\mathbf{r}) = z-h(\mathbf{r}_\perp).
\end{equation}
Inserting this expression into Eq.~\eqref{eq:boltzmann} and keeping only terms up to first order in $h$ results in
\begin{equation}
    P(\phi(\mathbf{r}))=\frac{1}{Z}\exp\left(-\frac{\frac{\kappa}{2}\int \mathrm{d}^3r \left[\epsilon f^{\prime}\left(\frac{z}{\sqrt{2}\epsilon}\right) \nabla^2 h(\boldsymbol{r}_\perp)\right]^2}{k_BT}\right).
\end{equation}
When evaluating the integration over $z$ in the limit $\epsilon\rightarrow 0$, we obtain
\begin{equation}
    P(\phi(\mathbf{r}))=\frac{1}{Z}\exp\left(-\frac{\kappa\frac{\sqrt{2}}{3}\int_S \mathrm{d}^2r_\perp \epsilon^3  \left(\nabla_\perp^2 h(\boldsymbol{r}_\perp)\right)^2}{k_BT}\right).
\end{equation}
From here, we can transform to Fourier space,
\begin{equation}
    h_\mathbf{k} = \int_S \mathrm{d}^2r_\perp h(\mathbf{r}_\perp)\exp(i\mathbf{kr}_\perp),
\end{equation}
to calculate the equilibrium height fluctuations
\begin{equation}
    \langle h_\mathbf{k} h_{-\mathbf{k}}\rangle = \frac{3 k_B T |S|}{2\sqrt{2}\kappa\epsilon^3 |\mathbf{k}|^4}.
    \label{eq:spectrum}
\end{equation}
$|S|$ is the surface area.
Thus, we recover the known result for the fluctuations of a planar membrane described by the free energy in Eq.~\eqref{eq:helfrich} (up to a constant factor)~\cite{seifertConfigurationsFluidMembranes1997}.
Our approach has the advantage that the fluctuation spectrum can be directly predicted from our microscopic equations.

\section{Numerical implementation}
We measure energies in units of $\kappa_\mathrm{eff}=\kappa\epsilon^3$, lengths in units of $\epsilon$, and times in units of $M^{-1}_\phi \kappa^{-1}$. 
This means that we effectively set $\epsilon=1$, $\kappa=1$, and $M_\phi=1$ in the above equations.
Typically, $k_BT$ is found in a range from $0.1\kappa_\mathrm{eff}$ to $0.01\kappa_\mathrm{eff}$ in experiments~\cite{monzelMeasuringShapeFluctuations2016}.
If not specified otherwise, we set $k_BT=0.033$, $\alpha=10$, and $\gamma=0.01$.
The dynamical equations are implemented using an Euler scheme, where the linear term in $\delta F_{bend}/\delta \phi$ containing $\nabla^4\phi$ is treated implicitly to increase stability.
The spatial derivatives are calculated using spectral methods.
We set the integration timestep to $\Delta t=0.1$ and $\nu=(\Delta t\Omega)^{-1}$, where $\Omega$ is the volume of the computational domain, to conserve $\int_\Omega \phi\ \mathrm{d}^3r$ exactly.
We discretize $\phi$ on a grid with grid constant $\Delta x = 1$.

\begin{figure}
    \includegraphics[width=.4\textwidth]{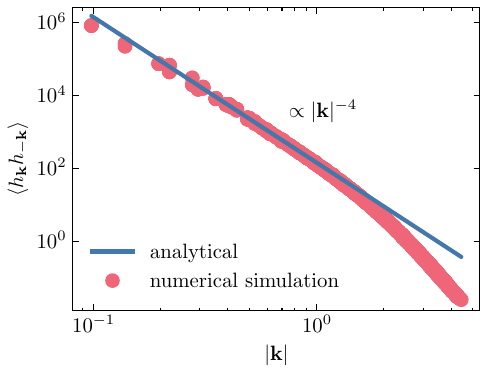}
    \caption{Fluctuations $\langle h_\mathbf{k}h_{-\mathbf{k}}\rangle$ of the height of the membrane around a flat quadratic patch of membrane of length $L=60$ at temperature $k_BT=0.033$ plotted over the wave number $|\mathbf{k}|$. The analytical expression in Eq.~\eqref{eq:spectrum} (blue line) is compared to the results from a simulation at $k_BT=0.033$ (red dots), averaged over a time of $t=10000$.
    For wavelengths above the width of the interface, $|\mathbf{k}|< \epsilon^{-1}$, the results from the analytical expression match those from the numerical simulation.}
    \label{fig:spectrum}
\end{figure}

To test the implementation, we have verified that the equipartition of energy is fulfilled, the dynamics therefore indeed results in the stationary probability distribution Eq.~\eqref{eq:prob}, see Appendix \ref{app:equ}.
We also measured the fluctuations of the membrane shape as predicted by Eq.~\eqref{eq:spectrum}.
For that purpose, we initialize a cubic computational domain with edge length 60 to contain two parallel flat layers of membrane at a distance of half the system size (an even number of layers is necessary to fulfill the periodic boundary conditions).
We measure explicitly the fluctuations in height by finding the zero-level-set surface via linear interpolation of the phase-field in the direction perpendicular to the initial surface.
From Fig.~\ref{fig:spectrum}, it can be inferred that the membrane fluctuations are indeed described by Eq.~\eqref{eq:spectrum}.

For the following measurements of diffusion through triply periodic membranes, the phase field describing the initial, unperturbed membrane is generated using the approach of Ref.~\onlinecite{yangPhasefieldApproachMinimizing2010}.
The computational domain consists of $3\times 3\times 3$ unit cells of the membrane, where one cell has a length of $d_c=40$.
A unit cell is therefore about $d_c/\sqrt{2}\approx 28$ times larger than the thickness of the transition layer of the phase field from one side of the membrane to the other.
This proportion is comparable to cubic membranes found in living systems where structures with unit cells of 50--550~nm can be formed by lipid bilayers with a thickness of less than 10~nm~\cite{almsherqiCubicMembranesMissing2009}.
For each set of parameters, we simulate 192 independent diffusion processes.
Since the presence of the diffusing object influences the membrane, we need to generate a separate realization of the membrane for each realization of the diffusion process.
For each realization, we evolve the membrane surface for at least $t_\mathrm{eq}=300000$, so that the shape fluctuations equilibrate.
This state is saved and used as a starting condition for the subsequent measurements.
We generate 64 such starting configurations.
The starting position of the diffusing object is in the center of a cavity of the membrane structure.
For each starting configuration of the membrane, we simulate three different starting positions of the particle.

\section{Results}
\begin{figure}
    \includegraphics[width=.4\textwidth]{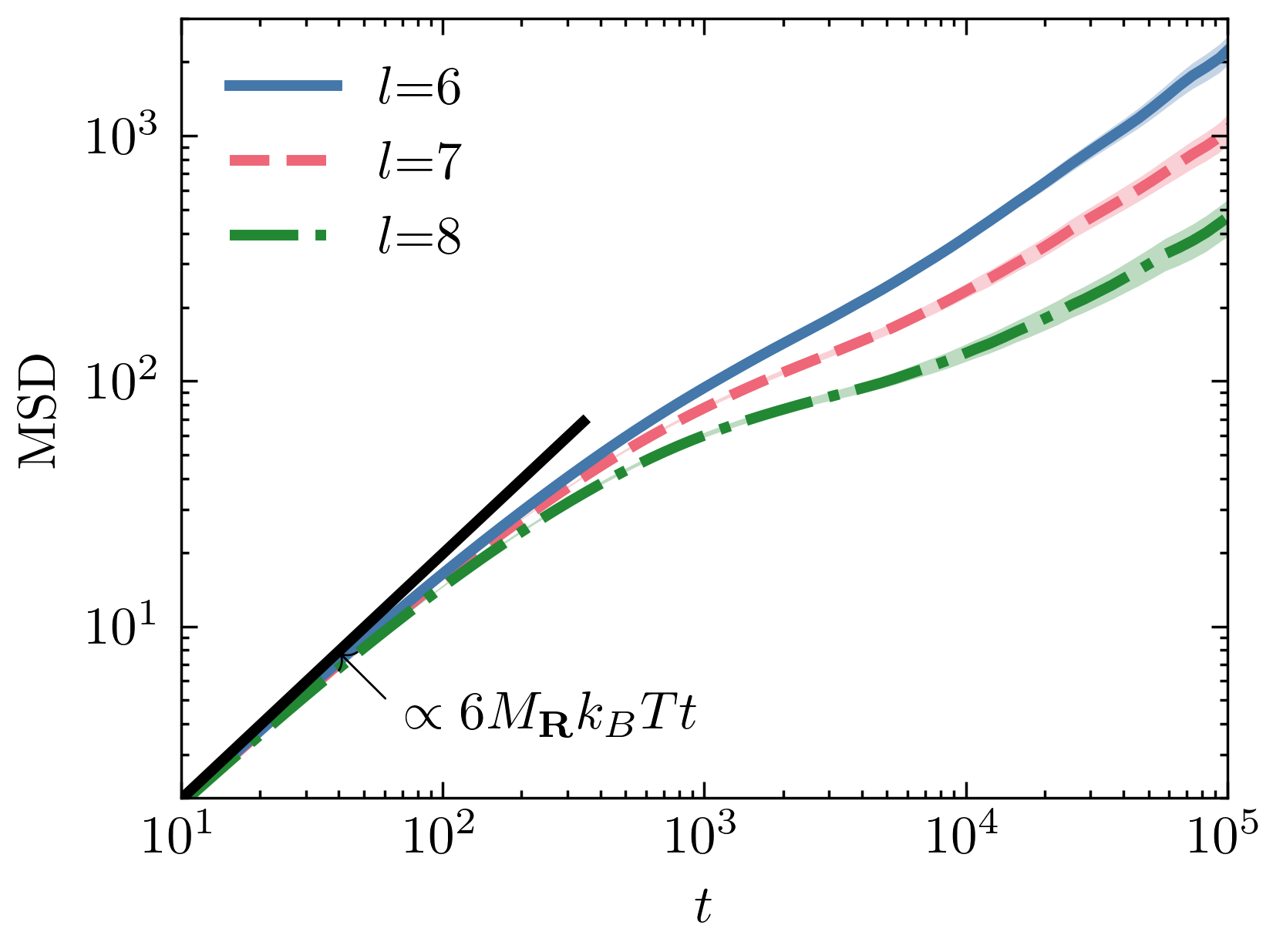}
    \caption{Averaged mean squared displacement (MSD) of the diffusing spherical object over the time $t$. The three different lines are for different particle radii $l$, while $M_\mathbf{R}=1$ is constant.
    Shaded areas indicate standard deviations of the means over 192 independent realizations. Two appropriately diffusive regimes of the MSD (black line) are separated by a region of reduced slope, which is caused by trapping in the cavities of the membrane structure. 
    With increasing particle radius, this region becomes more pronounced, for $l=8$ it is rather plateau-like.
    }
    \label{fig:MSD}
\end{figure}
\begin{figure}
    \includegraphics[width=.4\textwidth]{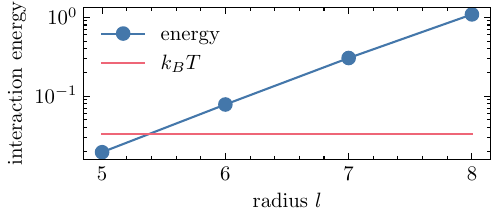}
    \caption{The energy barrier at the narrowest point of a channel for different particle radii $l$, calculated for the static, initial surface.
    For a unit cell size of 40, the narrowest channel formed by a P-surface has a radius of 10. However, due to the additional width of the transition layer, which is of order $\sqrt{2}$ for both the membrane and the particle, the effective radius is between 5 and 6.
    }
    \label{fig:barrier}
\end{figure}

We investigate diffusion through a P-surface as shown in Fig.~\ref{fig:example}.
First we keep $M_\mathbf{R}$ constant, while varying $l$. A larger radius of the diffusing spherical object makes it more difficult to fit through narrow pores.
Figure~\ref{fig:MSD} shows the mean squared displacement (MSD) for the three different radii of the diffusing particle, $l=6,7,8$, with $M_\mathbf{R}=1$ ($M_\mathbf{R}$ has the dimension of $M_\phi \epsilon^2$).
There are three different regimes. For short times, the MSD is linear with the diffusion coefficient of a free particle, $D_0=M_\mathbf{R}k_BT$.
At intermediate times, there is a subdiffusive regime, where the confinement by the membrane structure becomes apparent.
This regime starts a bit earlier for larger radii, because for larger particles the accessible volume inside a cavity is effectively smaller.
Similarly, the decrease in slope is more pronounced and resembles a plateau in Fig.~\ref{fig:MSD} for larger radii.
After some time, the particle escapes through a fluctuating pore into the next cavity.
The larger the particle is, the longer it takes until this escape happens.
We note that without the membrane fluctuations, this escape were not possible.
Although the narrowest point of a channel confined by the P-surface with a unit cell size of 40 has an approximate radius of 10, which is larger than all radii considered here, we need to take into account the transition layer of the phase-field describing the membrane $\phi$ and the particle $k(r)$.
From Eq.~\eqref{eq:tanh} and Eq.~\eqref{eq:kernel}, we know that it is approximately $\sqrt{2}$ wide but not sharply defined.
Thus, in order to determine the effective size of the pores exactly, we instead calculate directly the energy barrier for the particle.
As Fig.~\ref{fig:barrier} shows, the energy barrier at the narrowest point of a channel is larger than $k_BT$ for the particle radii considered in Fig.~\ref{fig:MSD}.
Finally, at longer times, the motion appears diffusive again, see Fig.~\ref{fig:MSD}.

\begin{figure}
    \includegraphics[width=.4\textwidth]{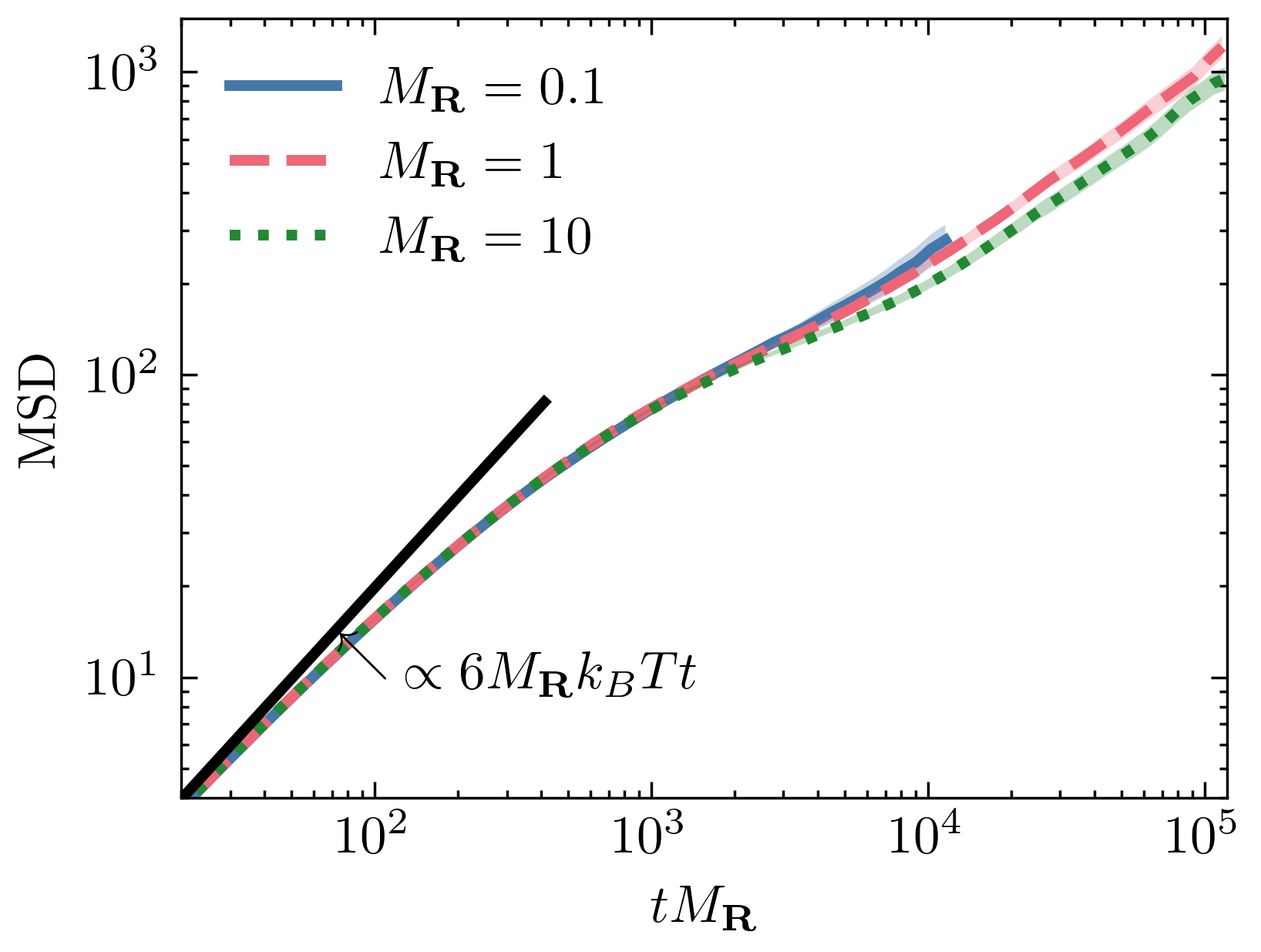}
    \caption{Averaged mean squared displacement (MSD) of the particle over the normalized time $tM_\mathbf{R}$. Shaded areas indicate standard deviations of the means over 192 independent realizations. The three different lines are for different mobilities $M_\mathbf{R}$, while $l=7$ is constant.}
    \label{fig:MSD2}
\end{figure}

For further analysis, we keep the radius $l$ constant and change the mobility $M_\mathbf{R}$.
Figure~\ref{fig:MSD2} shows the resulting MSD.
The time axis in this figure displays the time normalized to the timescale of free diffusion, $tM_\mathbf{R}$, so that we can compare measurements at different particle speeds.
We note that the MSD on small timescales therefore collapses on one common curve for all particle radii, as it is equal to that of a free particle.
However, when the dynamics of the particle is much faster ($M_\mathbf{R}=10$) than the dynamics of the membrane, the long-term MSD is lower than when the two timescales coincide ($M_\mathbf{R}=1$) or when the particle is slower ($M_\mathbf{R}=0.1$).
We suspect that this effect is related to the rate of jumping to the next cavity.
As an underlying reason, widening of pores due to thermal fluctuations makes slipping through a pore more likely.
For $M_\mathbf{R}=10$, the membrane becomes effectively static from the point of view of the particle, so that no new pathways can open. 
In the other extreme of $M_\mathbf{R}=0.1$, deformations of the membrane relax quickly compared to the speed of the particle.

Finally, we wish to consider the role of another potential effect that may facilitate diffusion of objects through pores that, in equilibrium, are of insufficient diameter. The object itself, when pushing through its motion into the pore, may deform the membrane and widen the pore to enable passage. 
This process goes beyond the fact that pores can open due to the fluctuations.

To investigate the relevance of these two different contributions in more detail, we introduce two variations of the dynamics for comparison.
In the first scenario the membrane is frozen in its deformed state and does not fluctuate at all. The motion of the particle is confined by this frozen membrane. 
In the second variation, the membrane is not affected by the presence of the particle at all. That is, there is no deformational effect of the particle on the membrane. However, the particle is still confined by the membrane.
Consequently, the system is no longer in equilibrium in both cases, as it no longer follows the stationary probability distribution Eq.~\eqref{eq:prob}.
Nevertheless, in our theoretical approach we can simply introduce these modifications and compare the results to conclude on the importance of the underlying processes. 

\begin{figure}
    \includegraphics[width=.4\textwidth]{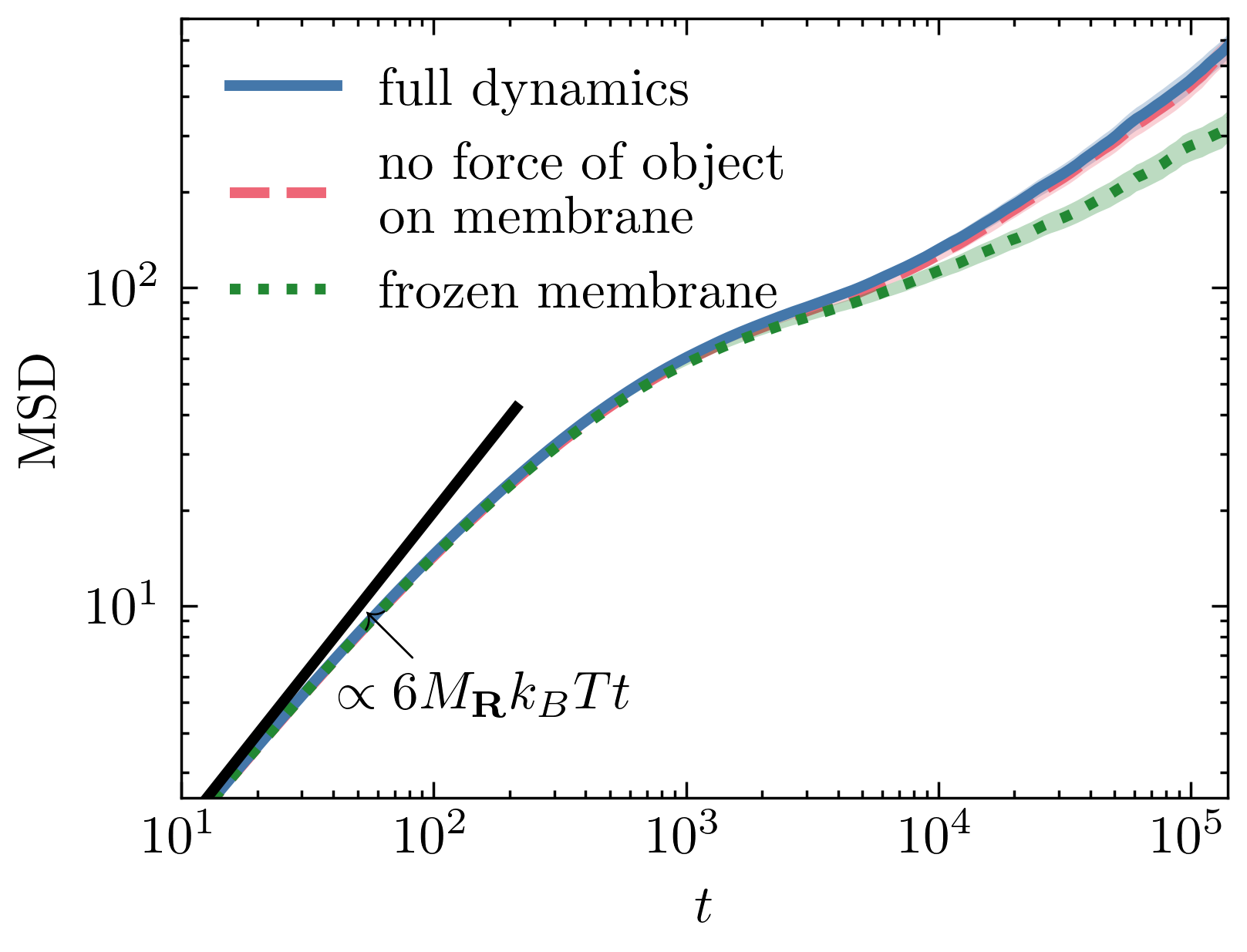}
    \caption{Averaged MSD of a diffusing particle with radius $l=8$. Shaded areas indicate the standard deviation of the mean over 192 independent realizations. The full dynamics are compared to a particle diffusing through a frozen membrane, and to a membrane that fluctuates but is not deformed by the particle.
    In the case of a frozen membrane, the long-time MSD is significantly decreased when compared to the full dynamics.
    }
    \label{fig:reason}
\end{figure}

In Fig.~\ref{fig:reason} we compare the MSD obtained from the two modifications to the unmodified equilibrium dynamics.
All other parameters are kept the same, including the random seeds for the fluctuations in our simulations.
In a frozen environment, we find that the particle moves slower than in a fluctuating environment.
This is in line with our previous results in Fig.~\ref{fig:MSD2}, as it essentially corresponds to the limit $M_\mathbf{R}/M_\phi\rightarrow\infty$.
Interestingly, when we deactivate the action of the particle on the membrane, we see no decrease in long-time diffusivity within our standard deviations.
This result confirms that the speed-up is significantly caused by the membrane fluctuations, and not notably by the force exerted by the particle on the membrane, which may contribute to widen the pores.

\begin{figure}
    \includegraphics[width=.25\textwidth]{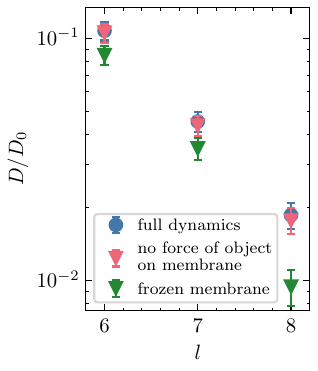}
    \caption{Long-time diffusion coefficient $D$, normalized by $D_0=M_\mathbf{R}k_BT$.
    Error bars show estimated standard deviations from the means as obtained from sampling 192 realizations.
    The full dynamics are compared to the scenario of suppressed deformational action of the particle on the membrane, with no notable deviation within the standard deviations. Moreover, we compare to a scenario of a frozen membrane.
    The latter situation leads to notably smaller diffusion coefficients, especially for large diffusing objects.}
    \label{fig:diffusivity}
\end{figure}
We extend this comparison to different radii of the particle. 
In Fig.~\ref{fig:diffusivity}, we show the long-time diffusion coefficient, defined by $D=\lim_{t\to\infty}\langle \mathbf{r}^2\rangle/6t$.
It is obtained from a least-squares fit to the MSD in the interval $t=20000$ to $t=100000$.
With increasing size of the diffusing object, the decrease of the diffusion coefficient for a frozen membrane becomes larger.
A conceivable reason refers to the confinement that is stronger for particles with larger radius, which increases the importance of fluctuation events to open the pores for passage.

\section{Conclusions}
We have investigated the motion of a diffusing object through fluctuating triply periodic membrane structures.
Our investigation was accomplished using a phase-field approach to describe the confinement of the object by the membrane, including fluctuations of the membrane.
We have shown analytically and numerically, that adding a stochastic term to equations of motion of the phase-field membrane indeed leads to the correct statistics of interface displacements.

In the mean squared displacement of the diffusing object, we identified two linear diffusive regimes that are separated by a plateau-like subdiffusive intermediate region that corresponds to the trapping of the particle inside a cavity of the periodic membrane structure.
This region becomes broader when the size of the object is increased.
The fluctuations of the membrane, particularly of the diameter of the pores, allow for the escape of a particle that would be trapped inside a given pore if the membrane surface were unperturbed.
This effect should be directly observable, for instance, in protein diffusion through biological environments.
Our theoretical approach allows us to separate the effect of the fluctuations of the membrane and pore size from the effect that the diffusing object has on the membrane.
When by hand we suppressed deformation of the membrane by the diffusing object, the long-time diffusion coefficient remained virtually unchanged.
We thus conclude that the dominant effect is the opening of the pores caused by fluctuations of the membrane, and not deformations of the soft membrane induced by the diffusing object.

In the future, the dynamics of the phase field and the object should be coupled to the Navier-Stokes equation of hydrodynamic fluid flow to further include the role of the liquid environment~\cite{campeloDynamicModelStationary2006}.
Long-range hydrodynamic interactions of the membrane with itself and with the diffusing object could be included in this way.
Such couplings may play an important role in transport through biological systems as well.

\section*{Acknowledgement}
The authors thank Gerd Schröder-Turk for a stimulating discussion on minimal surfaces at the beginning of this study.

\bibliography{tpms}
\pagenumbering{arabic}

\appendix
\section{Test of the numerical implementation}
\label{app:equ}
\begin{figure}[h]
    \includegraphics[width=.4\textwidth]{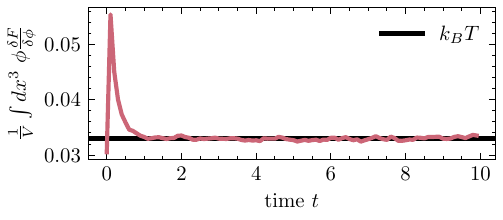}
    \caption{A spatial average of $\langle\phi{\delta F}/{\delta \phi}\rangle$ (red line) that quickly approaches $k_BT$ (black line) as a function of simulation time after initialization at $t=0$.
    Parameters are $k_BT=0.033$ and $\Delta x =1$. The system was initialized to $3\times 3\times 3$ unit cells of the P-surface, where one cell has a length of 40.}
    \label{fig:thermalization}
\end{figure}

We test the implementation first by measuring $\phi{\delta F}/{\delta \phi}$.
If the system is described by the probability distribution in Eq.~\eqref{eq:prob}, equipartition dictates
\begin{equation}
    \left\langle\phi\frac{\delta F}{\delta \phi}\right\rangle = k_BT.
\end{equation}
Figure~\ref{fig:thermalization} shows that this is indeed fulfilled.

\end{document}